\documentclass[aps,prl,twocolumn]{revtex4}
\usepackage{epsfig}

\begin{document}

\title{Orange Peel coupling in granular ferromagnetic films}
\author{D. Barness}
\author{A. Frydman}
\address{Department of Physics, Bar Ilan University, Ramat Gan 52900, Israel}

\begin{abstract}
We present magnetoresistance (MR) measurements performed on
magnetic tunnel junctions in which one of the electrodes is a
granular ferromagnetic film. These junctions exhibit a zero field
resistance dip. The dip magnitude depends on the size of the
grains. We interpret these results as a consequence of the orange
peel effect between the continuous ferromagnetic film and the
magnetic grains. The coupling is found to be much stronger than
that between continuous ferromagnetic layers.

\end{abstract}

\newpage
\maketitle GMR (giant magnetoresistance) and TMR (tunneling
magnetoresistance) devices are primary candidates in future
magneto-electronic applications and media \cite{baibich, chou,
prinz, vouille, fert}. The ability to create arrays of magnetic
junctions on micro sized areas can enhance storage size
drastically and enable the production of non volatile ultra-dense
RAM chips. Granular ferromagnets have a promising potential to act
as a further step in this direction since a single junction may be
able support numerous bits, considerably increasing the possible
storage densities.

All applications based on GMR and TMR effects require high quality
multilayers constructed of thin ferromagnetic and non-magnetic
films. The performance of the devices depend strongly on the
morphological and structural properties of the films as well as
their physical characteristics. Among the crucial factors is the
interlayer coupling between two ferromagnetic layers separated by
a non-magnetic spacer. This coupling may be a sum of several
different mechanisms, of which three appear to be dominant. The
first is pinhole coupling, which results from structural defects
in the spacer and may destroy MR effects altogether. The second is
the RKKY interaction which oscillates with the spacer thickness.
This coupling is due to indirect exchange mechanism and applies
only to conductive barriers (GMR multilayers). The third mechanism
is the N\'{e}el Coupling \cite{Neel}, also named Orange Peel
Effect (OPE), which applies both to conducting and to insulating
spacers. This coupling utilizes the surface waviness of correlated
layers to produce ferromagnetic interaction between ferromagnetic
layers that could otherwise be antiferromagnetically coupled. The
mechanism is based on the fact that the waviness of the magnetic
film creates dipoles on the surface. A second layer with
correlated waviness placed on top and separated by a non-magnetic
spacer, experiences similar moment orientation due to
dipole-dipole interaction as illustrated in figure 1a. Such
ferromagnetic coupling reduces the GMR signal which requires
antiferromagnetic orientation at low fields. Hence, a lot of
effort is invested in an attempt to minimize this coupling in
order to improve the performance of GMR/TMR elements \cite{chopra,
portier, shrag, tegen, koolsexp, wang}.

The basic N\'{e}el model was derived for two infinitely thick magnetic
layers separated by a non magnetic spacer \cite{Neel}. Kools et al \cite%
{kools} extended the theory to included the finite size of the magnetic
layers and obtained the following expression for the coupling strength:

\begin{equation}
H_{N\acute{e}el}=\frac{\pi ^{2}h^{2}M_{P}}{\sqrt{2}\lambda t_{F}}\left[
1-\exp \left( \frac{-2\pi \sqrt{2}t_{P}}{\lambda }\right) \right] \exp
\left( \frac{-2\pi \sqrt{2}t_{s}}{\lambda }\right)   \label{neel}
\end{equation}

where h and $\lambda $ are the amplitude and wavelength of the layer
waviness. $t_{F}$, $t_{p}$ and $t_{s}$ are the thicknesses of the free
layer, the pinned layer and the spacer respectively, and $M_{P}$ is the
magnetization of the pinned layer.

So far, N\'{e}el coupling was studied only between continuous
ferromagnetic layers. One may ask whether a similar effect can
take place between a ferromagnetic layer and a set of
ferromagnetic grains. This issue can be of great importance for
the design of magnetic devices based on granular structures.
A-priori there seems to be no reason why OPE should not apply to
granular systems providing the size of the grains correlates with
the surface roughness of the layer. A schematic description of
such a possible coupling is depicted in figure 1b. In this paper
we describe an experimental effort to explore the OPE in granular
systems. We present magnetoresistance measurements of magnetic
tunnel junctions consisting of a uniform Ni layer, an insulating
spacer and a granular Ni film. These structures exhibit a sharp
magnetoresistance dip at low magnetic fields which are interpreted
as signs for the OPE in the single domain grains.

\begin{figure}\centering
\epsfxsize7.5cm\epsfbox{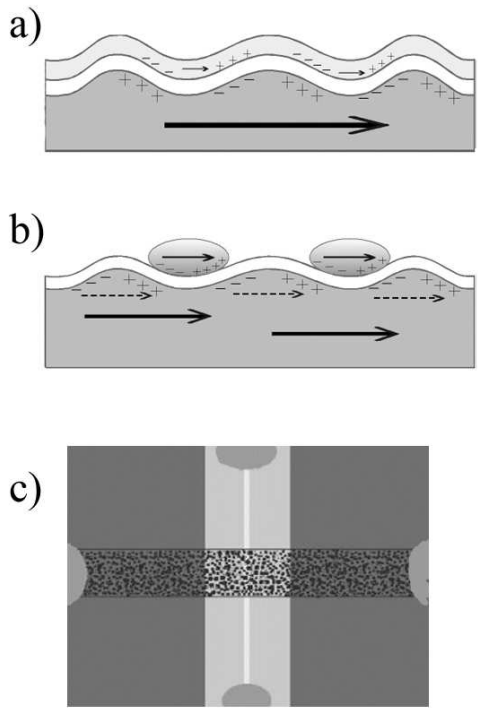} \caption{a) Schematic
description of the orange peel effect model for two finite
continuous layers. Poles created on the surface of the bottom
layer (pinned) induce compatible moments in the upper layer
(free), thus creating artificial magnetic orientation. b) The
possible orange peel model with a granular layer functioning as
the free layer. The grain size is of the order of the waviness.
Here the poles induced by the pinned layer create artificial
magnetic orientations on individual grains. c) A sketch of the
junction geometry. } \label{fig1}
\end{figure}

The samples were prepared using the following scheme: First, a
250\AA\ thick Ni layer was e-beam evaporated on a Si/SiO substrate
through a 1 mm wide mask. The deposition rate was 0.5 \AA /s and
the base vacuum was 10$^{-7}$ Torr. Next, a 100\AA\ layer of SiO
was deposited in a 10$^{-4}$ mbar oxygen environment through a
0.25 mm wire-shape window leaving a narrow slit on top of the Ni
layer. A 30\AA\ thick Al$_{2}$O$_{3}$ (the insulating spacer) was
then deposited into the window in a 10$^{-4}$ mbar O$_{2}$
environment and at a rate of 0.2 \AA /s. The quality of the
barrier was confirmed by measuring the I-V curves of similar based
Ni/Al$_{2}$O$_{3}$/Pb superconducting junctions. Finally a
discontinuous Ni film was grown on top through a narrow strip
shadow mask vertical to the continuous Ni layer, thus completing a
4 terminal junction geometry of 0.2 mm$^{2}$ as illustrated in
figure 1c.

A major concern with the granular system is the prevention of structural and
chemical changes of the film due to oxidation or annealing. In order to
circumvent these problems we prepared the granular films using the technique
of quench condensation, i.e. evaporation on a cryo-cooled substrate within
the measurement probe under ultra high vacuum conditions \cite%
{strongin,granular bob,granular goldman, granular rich}. This
technique enables the growth of ultra-clean stable Ni grains with
an excellent control over the inter-grain distance and the film
resistance \cite{aviad1, aviad2}. For our junctions we quench
condensed Ni discontinuous film on a substrate held at T=5K and
with deposition rate ranging between 0.03-1\AA\/s. As will be
shown later the average grain size depends upon this rate. The
sample resistance and thickness were measured during growth and
the process was stopped at a desired resistance of 30-300K$\Omega
$. This resistance range insured that the film had an insulating
granular geometry but, on the other
hand, its resistance was much smaller than that of the junction (a few M$%
\Omega )$, thus ensuring that the granular layer can be regarded as an
equipotential electrode.

MR studies were performed at T=4K with external field applied
perpendicular to the junction's plane. All measurements were
performed using standard lock-in AC techniques and making sure
that the I-V curves were in the ohmic regime. Figure 2 depicts the
MR curve of 3 typical junctions in which the granular film was
evaporated at rates of 1,0.3 and 0.03 \AA /s. It is seen that the
samples exhibit the following main features: As the field is
reduced from high fields to below $\pm $1.2T a resistance rise of
2-15\% is observed. This rise is rather gradual and extends up to
about 0.4T. Further reducing the field below $\pm $\ 0.2T causes a
sharp resistance drop so that the MR curve exhibits a resistance
dip around H=0. We note that all of our measured samples exhibited
these features. However, the magnitudes of the low field
resistance minimum as well as the high field features depended
strongly on the deposition rate of the granular film.

\begin{figure}\centering
\epsfxsize7.5cm\epsfbox{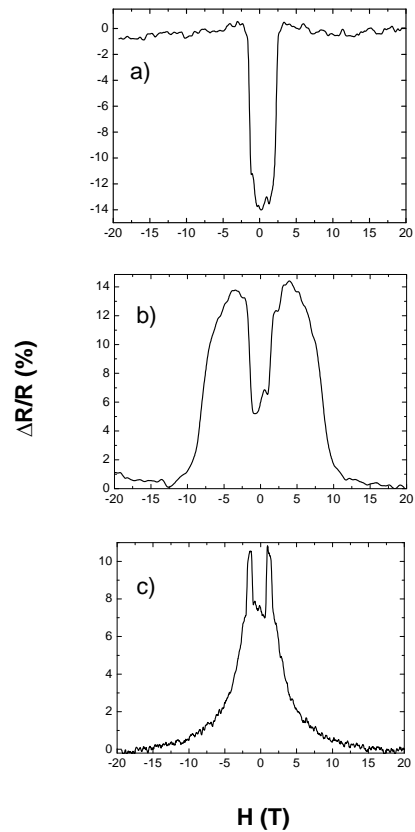} \caption{MR curves of three
typical tunnel junctions in which the granular films were
deposited at different rates: 1, 0.3 and 0.03 \AA/s, for a) b) and
c) respectively. Notice the difference in the magnitude of the
zero field resistance dip. The faster the evaporation rate the
larger the dip.} \label{fig2}
\end{figure}

The high field resistance increase with decreasing magnetic field
can be expected. The grain anisotropy may be different than that
of the continuous film, hence, the grains may switch their
magnetic orientation at fields much larger than that of the
continuous film. This leads to a resistance increase. We note that
the fact that this feature extends to relatively large fields is
unusual for Ni where all elements are expected to reach saturation
at fields of $H\cong 0.6T.$ In a different study we suggest
\cite{doron} that these unusually high fields are a result of the
existence of ultra small grains that experience unusual high
coercivity.

The sharp resistance dip at small fields is more surprising. For
this field range the magnetic elements of the junction are
expected to be randomly oriented and the resistance is expected to
exhibit a \textit{maximum}. The sharp \textit{minimum} at H=0 is
indicative of magnetic alignment of the continuous film with the
grains on top. This requires a mechanism for low field
ferromagnetic coupling between the uniform Ni film and the
granular Ni system. A natural candidate is the Orange Peel effect.
As noted above, this effect was extensively studied for systems
containing two ferromagnetic layers. If a similar mechanism is
relevant in a multilayer containing a granular film, the thickness
roughness of the continuous film has to correlate with the grain
average size. Figure 3 shows AFM analysis of a continuous Ni film
and of granular films deposited at different deposition rates. It
is important to note that it is very difficult to obtain exact
morphological information on the quench-condensed granular films.
As the system is heated to room temperature the films may
experience some degree of creep and grains may slightly move.
However, the AFM pictures can provide a good approximation for the
structural features of the system. Figure 3 illustrates that the
thickness roughness of our continuous layers has a wavelength of
about 400-500 \AA . For the granular systems, the average grain
size grows as a function of deposition rate and approaches the
wavelength of the continuous layer roughness for the fastest
deposited films. A comparison between figures 2 and 3 reveals a
clear correlation between the size of the grains and the amplitude
of the zero bias dip in figure 2. The larger the grains (and the
closer their size to the scale of the roughness in the continuous
layer) the larger the dip. Granular systems which were grown at a
rate of 1 \AA /s have an average size of 400 \AA\ and exhibit a
dip magnitude of about 15\% (figures 2a and 3a). For a deposition
rate of 0.3 \AA /s we get grain sizes of about 320 \AA\ and a dip
of 8 \% (figures 2b and 3b) and for a rate of 0.03 \AA /s the
sizes are 250 \AA\ and the dip is 4 \%(figures 2c and 3c).

\begin{figure}\centering
\epsfxsize7.5cm\epsfbox{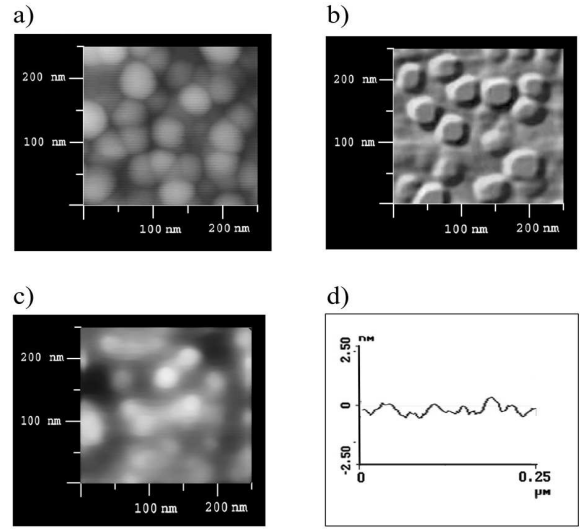}\vskip -3truecm \caption{AFM
analysis of our layers. a b and c are AFM images of granular Ni
films evaporated at rates of 1, 0.3 and 0.03 \AA/s respectively
(corresponding to the samples of figure 2a, 2b and 2c). d is the
line scan of a continuous 250 \AA  thick  Ni film.} \label{fig3}
\end{figure}

This correlation is in agreement with our notion that the resistance dip is
due to N\'{e}el coupling between the continuous layer and grains on top. We
envision that at very high fields both the continuous and the granular films
are oriented parallel to the field direction and the resistance is low. As
the field is reduced magnetic moments in the smallest grains start falling
into the plain thus increasing the resistance. This process continues down
to fields of the order of 0.4T in which the magnetic moments of the
continuous film start spraying into the plain. At fields below 0.2T the OPE
becomes important and each grain moment aligns itself with the moment of the
domain placed below. This leads to the resistance dip at H=0.

A feature which remains peculiar is the fact that for the high
deposition rate samples (figure 2a) the resistance at H=0 may be
smaller than the high field value. One would expect that at high
fields all elements are aligned and the resistance should be
minimal. We note that such anomalies were seen in very dilute
granular films \cite{amitay}. In these samples the high field
resistance was often higher than the zero field state . This
effect was ascribed to magnetic dipole-dipole repulsion that the
grains experience when they are aligned perpendicular to the
substrate. We propose similar considerations in our junction
geometry. As grains start aligning parallel to each other, all
perpendicular to the substrate, dipole-dipole interactions cause
magnetic repulsion between the grain moments which are aligned
parallel to each other. This repulsion prevents the dipoles from
being fully aligned and therefore the resistance of the junction
is not at its minimum at high fields. At low fields the OPE aligns
the grains parallel to the bottom continuous layer and the above
reasoning is not relevant, hence a lower tunneling resistance may
occur.
If indeed the cause for the resistance dip is the OPE we
can estimate its strength using equation 1. Based on our AFM
images we use the following geometrical values: h=20\AA , $\lambda
=400$\AA\ , t$_{F}=25$\AA , t$_{P}=250$\AA\ and t$_{S}=25$\AA . Mp
for Ni is 560 esu/cc to yield a coupling strength of
H$_{N\acute{e}el}\approx 60Oe$. This value is similar to coupling
strengths estimated and measured in continuous layer structures in
which the OPE was studied. Our experimental results, however show
that
the coupling field scale extends to magnetic fields of approximately \textit{%
2000 Oe}. This large coupling scale seems to be a unique feature
of our granular systems. Figure 2 demonstrates that although the
magnitude of the dip depends on the deposition rate of the grains,
the field scale of the effect is similar for all granular systems.
This peculiar feature requires further theoretical treatment.

In conclusion we have measured the magnetoresistance properties of a
multilayer containing continuous and granular Ni films separated by an
insulating barrier. These exhibit a low field resistance dip which is
indicative of a ferromagnetic coupling between the films. The coupling
becomes stronger, the larger the correlation between the grain size and the
surface roughness of the continuous film. We interpret these findings as
signs for the N\'{e}el coupling taking place between the continuous film and
the grains. The coupling seems to be much stronger than that between two
continuous layers.

We gratefully acknowledge illuminating discussions with F. Hellman
and technical help from A. Cohen. This research was supported by
the Israel Science foundation grant 326/02.

\newpage

\end{document}